\documentclass[aps,pra,onecolumn,showpacs,superscriptaddress]{revtex4}
\usepackage{bm}
\usepackage{mathrsfs}
\usepackage{amsmath}
\usepackage{amssymb}
\usepackage{graphicx}
\usepackage{amsfonts}
\usepackage{amsthm}
\usepackage{color}
\usepackage{dcolumn}
\usepackage{txfonts}
\usepackage{epsfig}

\begin{document}
\title{Fluctuation Relations for Heat Exchange in the Generalized Gibbs Ensemble}
\author{Bo-Bo Wei}
\email{Corresponding author: bbwei@szu.edu.cn}
\affiliation{School of Physics and Energy, Shenzhen University, Shenzhen 518060, China}

\begin{abstract}
In this work, we investigate the heat exchange between two quantum systems whose initial equilibrium states are described by the generalized Gibbs ensemble.
First, we generalize the fluctuation relations for heat exchange discovered by Jarzynski and W\'{o}jcik to quantum systems prepared in the equilibrium states described by the generalized Gibbs ensemble 
at different generalized temperatures.
Second, we extend the connections between heat exchange and R\'{e}nyi divergences to quantum systems with very general initial conditions.
These relations are applicable for quantum systems with conserved quantities and are universally valid for quantum systems in the integrable and chaotic regimes.
\end{abstract}
\maketitle

\section{Introduction}
In 2004, Jarzynski and W\'{o}jcik found that for two quantum systems initially prepared at their thermodynamic equilibrium states with different temperatures $T_A$ and $T_B$ respectively the heat exchange $Q$
between them satisfies the fluctuation theorem \cite{Jarzynski2004}
\begin{eqnarray}\label{fa}
\frac{P(Q)}{P(-Q)}=e^{\Delta\beta Q},
\end{eqnarray}
where $P(Q)$ is the distribution of heat exchange and $\Delta\beta\equiv\beta_B-\beta_A$ with $\beta=1/T$ being the inverse temperature.
Recently, the author found that the heat exchange $Q$ between two quantum systems is related to the R\'{e}nyi divergences between the initial equilibrium state of the total system
and the final non-equilibrium state of the total system by \cite{Wei2017d}
\begin{eqnarray}\label{fb}
\left\langle \Big(e^{-\Delta\beta Q}\Big)^z\right\rangle =e^{(z-1)S_{z}[\rho(0)||\rho(\tau)]},
\end{eqnarray}
where $z$ is an arbitrary real number, $\Delta\beta\equiv\beta_B-\beta_A$ with $\beta=1/T$ being the inverse temperature and the angular bracket on the left side of Equation \eqref{fb} means
average over ensemble repetitions of initial equilibrium state and the order-$z$ R\'{e}nyi divergence between
$\rho(0)$ and $\rho(\tau)$, which are respectively the initial equilibrium state of the total system and the final non-equilibrium state of the total system at time $\tau$,
is defined by $S_{z}[\rho(0)||\rho(\tau)]\equiv\frac{1}{z-1}\ln[\text{Tr}[\rho(0)^{z}\rho(\tau)^{1-z}]]$ \cite{Renyi1961,Erven2014,Beigi2013,Lennert2013}.
Equation \eqref{fb} relates the heat exchange between two systems and the R\'{e}nyi divergences between microscopic states and thus various moments of the heat exchange are quantified by the relative entropy and R\'{e}nyi divergences
between microscopic states \cite{Wei2017d}. The details of various exchange fluctuation relations could be found in \cite{Jar2011,RMP2011}.

In the exchange fluctuation relations, Equation \eqref{fa} and \eqref{fb}, one assumes that both systems are initially prepared at their own thermodynamic equilibrium states at different temperatures
described by the Gibbs ensemble \cite{Jarzynski2004,RMP2011}. However, It is known that the equilibrium state of a number of quantum systems can not be described by the Gibbs ensemble \cite{GGE0,GGE1,GGE2,GGE3,GGE4,GGE5,GGE6,GGE7}.
For example, the quantum integrable systems possess a number of conserved quantities which constraint the quantum dynamics of the integrable system \cite{GGE0,GGE1,GGE2,GGE3,GGE4,GGE5,GGE6,GGE7}.
The steady state of a quantum integrable system can not be described by the Gibbs ensemble but are characterized by the generalized Gibbs ensemble \cite{GGE1,GGEreview2016,GGEreview2016b}.
Recently the generalized Gibbs ensemble has been tested experimentally \cite{GGEexp}.
The purpose of this work is to investigate the heat exchange between two quantum systems whose initial states are best described by the generalized Gibbs ensemble.
We found that the heat exchange between quantum systems described by the generalized Gibbs ensemble satisfies heat exchange fluctuation relations of the form in Equation \eqref{fa}
if the generalized heat exchange are defined. Moreover, the heat exchange in the generalized Gibbs ensemble is determined by the R\'{e}nyi divergences between microscopic states.

The paper is structured as follows: In Sec. II, we review some fundamentals of the generalized Gibbs ensemble.
In Sec. III, we study heat exchange between quantum systems with conserved quantities and then derive the fluctuation relations
for heat exchange between quantum systems whose initial equilibrium states are described by the generalized Gibbs ensemble.
In Sec.~IV, the relations between heat exchange and the R\'{e}nyi divergences are established in the generalized Gibbs ensemble. In Sec. V, we make a conclusion.

\section{The Generalized Gibbs Ensemble}
Ensemble theory is fundamental to the equilibrium thermodynamics and statistical physics \cite{Reichl1987}.
It is known that the statistical ensembles can be derived by the principle of entropy maximization pioneered by Jaynes \cite{Jaynes1957a,Jaynes1957b}.
For instance, the Gibbs ensemble theory may be deduced by the maximization of the entropy under the constraint that the average energy of the system is fixed
and the grand canonical ensemble theory can be derived by the entropy maximization under the constraints that the average energy and the average number of particles are fixed in the equilibrium.
However, an important number of physical systems have additional conserved quantities except the energy and particle number.
The equilibrium state for quantum system with conserved quantities may also be formulated by the principle of entropy maximization
and it turns out that \cite{GGE1} the generalized Gibbs ensemble is established to describe the equilibrium state of the quantum systems with more conserved quantities.
For a quantum system with Hamiltonian $H$ and $M$ additional conserved quantities $\{I_1,I_2,\cdots,I_M\}$, the equilibrium state in the generalized Gibbs ensemble is \cite{GGE1}
\begin{eqnarray}
\rho_{\text{GGE}}&=&\frac{1}{Z}\exp\left[-\left(\beta H+\sum_{k=1}^M\beta_kI_k\right)\right],
\end{eqnarray}
where $Z(\beta,\beta_1,\cdots,\beta_M)=\text{Tr}\left[\exp\left(-\beta H-\sum_{k=1}^M\beta_kI_k\right)\right]$ is the partition function of the generalized Gibbs ensemble.
We assume all conserved quantities commute with the Hamiltonian and the conserved quantities also commute with each other in order that they can be measured simultaneously.
The generalized inverse temperatures $\{\beta,\beta_1,\cdots,\beta_M \}$ may be determined by imposing that the generalized Gibbs ensemble averages give the known initial values of the energy $\bar{E}$ and other conserved quantities $\bar{I}_1,\bar{I}_2,\cdots,\bar{I}_M$,
\begin{eqnarray}
\langle H\rangle&=&\text{Tr}[\rho_{GGE}H]=\bar{E},\\
\langle I_k\rangle&=&\text{Tr}[\rho_{GGE}I_k]=\bar{I}_k, k=1,2,\cdots, M.
\end{eqnarray}
Recently the generalized Gibbs ensemble has been tested in experiments \cite{GGEexp}.  In the present work, we investigate the heat exchange between quantum systems whose initial states are described by the generalized Gibbs ensemble (GGE).

\section{Heat Exchange in the Generalized Gibbs Ensemble}
In this section, we investigate the fluctuation relations for heat exchange between quantum systems which are initialized in the generalized Gibbs ensemble (GGE).

\subsection{Distribution of Heat Exchange in Quantum Systems Described by GGE}
\noindent The heat exchange process between two quantum systems $A$ and $B$ which are initialized in the generalized Gibbs ensemble can be described by the following steps:\\
1. Consider the two systems $A$ and $B$ and we assume both of them are integrable systems and thus their equilibrium states are described by GGE. We assume the system $A$ has the following conserved quantities
\begin{eqnarray}
H_A,I_{1}^A,I_{2}^A,\cdots I_M^A,J_1,J_2,\cdots,J_{M'}.
\end{eqnarray}
Here $H_A$ is the Hamiltonian of the system $A$ and the total number of conserved quantities in system $A$ is $M+M'+1$. While the system $B$ has conserved quantities,
\begin{eqnarray}
H_B,I_{1}^B,I_{2}^B,\cdots I_M^B,K_1,K_2,K_3,\cdots,K_{N}.
\end{eqnarray}
Here $H_B$ is the Hamiltonian of the system $B$ and the number of conserved quantities in system $B$ is $M+N+1$. Note that the conserved quantities $I_1,\cdots,I_M$ are common to both systems while $J_1,J_2,\cdots,J_{M'}$ are
conserved quantities only in system $A$ and $K_1,K_2,\cdots,K_N$ are conserved quantities only in system $B$. The two systems are initialized in their thermodynamics equilibrium states and the initial state of the total system
is given by
\begin{eqnarray}
\rho(0)&=&\frac{e^{-(\mathcal{H}_A+\mathcal{H}_B)}}{Z_AZ_B},
\end{eqnarray}
where we define $\mathcal{H}_A\equiv\beta_0^A H_A+\sum_{i=1}^M\beta_{i}^AI_{i}^A+\sum_{i=1}^{M'}\lambda_{i}J_{i}$ and $\mathcal{H}_B\equiv\beta_0^B H_B+\sum_{i=1}^M\beta_{i}^BI_{i}^B+\sum_{i=1}^{N}\alpha_{i}K_{i}$,
$Z_A=\text{Tr}[e^{-\mathcal{H}_A}]$ and $Z_B=\text{Tr}[e^{-\mathcal{H}_B}]$ are respectively the partition functions in the GGE for systems $A$ and $B$ respectively.
The equilibrium state of system $A$ is characterized by the generalized temperatures $\vec{\beta}_A=(\beta_0^A,\beta_1^A,\cdots,\beta_M^A)$ and $\vec{\lambda}=(\lambda_1,\lambda_2,\cdots,\lambda_{M'})$.
The GGE state of $B$ is captured by the generalized temperatures $\vec{\beta}_B=(\beta_0^B,\beta_1^B,\cdots,\beta_M^B)$ and $\vec{\alpha}=(\alpha_1,\alpha_2,\cdots,\alpha_{N})$.
Here we make use of the vector notation to simplify the expressions.\\
2. At $t=0$, we measure the energy and other conserved quantities of the system $A$, $H_A,I_{1}^A,I_{2}^A,\cdots I_M^A,J_1,J_2,J_3,\cdots,J_{M'}$
and that of the system $B$, $H_B,I_{1}^B,I_{2}^B,\cdots I_M^B,K_1,K_2,K_3,\cdots,K_{N}$ respectively and the outcomes of the first projective measurement are
$E_{n}^A,I_{1,n}^A,I_{2,n}^A,\cdots,J_{1,n},J_{2,n},\cdots J_{M',n}$ in system $A$ and $E_{n}^B,I_{1,n}^B,I_{2,n}^B,\cdots,K_{1,n},K_{2,n},\cdots K_{N,n}$ in system B respectively.
The measurement results are the eigenvalues of the corresponding operators
\begin{eqnarray}
H_A|n\rangle&=&E_n^A|n\rangle,\\
I_i^A|n\rangle&=&I_{i,n}^A|n\rangle, i=1,2,\cdots, M,\\
J_i|n\rangle&=&J_{i,n}|n\rangle, i=1,2,\cdots,M',\\
H_B|n\rangle&=&E_n^B|n\rangle,\\
I_i^B|n\rangle&=&I_{i,n}^B|n\rangle, i=1,2,\cdots, M,\\
K_i|n\rangle&=&K_{i,n}|n\rangle, i=1,2,\cdots,N.
\end{eqnarray}
To simplify the notation, we define the following vectors,
\begin{eqnarray}
\vec{\mathcal{E}}_{n}^A&=&\left(E_{n}^A,I_{1,n}^A,I_{2,n}^A,\cdots,I_{M,n}^A\right),\\
\vec{\mathcal{E}}_{n}^B&=&\left(E_{n}^B,I_{1,n}^B,I_{2,n}^B,\cdots,I_{M,n}^B\right),\\
\vec{\mathcal{J}}_{n}&=&\left(J_{1,n},J_{2,n},\cdots J_{M',n}\right),\\
\vec{\mathcal{K}}_{n}&=&\left(K_{1,n},K_{2,n},\cdots K_{N,n}\right).
\end{eqnarray}
With this shorthand notation, the outcome of the first projective measurement is $\{\vec{\mathcal{E}}_{n}^A,\vec{\mathcal{E}}_{n}^B,\vec{\mathcal{J}}_{n},\vec{\mathcal{K}}_{n}\}$ with the corresponding probability
\begin{eqnarray}
p_n(0)=\frac{e^{-\left(\vec{\beta}_A\cdot\vec{\mathcal{E}}_{n}^A+\vec{\lambda}\cdot\vec{\mathcal{J}}_{n}+\vec{\beta}_B\cdot\vec{\mathcal{E}}_{n}^B+\vec{\alpha}\cdot\vec{\mathcal{K}}_{n}\right)}}{Z_AZ_B}.
\end{eqnarray}
Simultaneously the quantum state of the total system becomes $|n\rangle=|n_A,n_B\rangle$. \\
3. We then let the systems $A$ and $B$ contact and interact for a time duration $\tau$. We assume the total Hamiltonian describing the interaction between $A$ and $B$ is
\begin{eqnarray}
\mathcal{H}=H_A+H_B+H_{AB}.
\end{eqnarray}
Then the quantum state of the total system at $t=\tau$ is $\mathcal{U}_{0,\tau}|n\rangle$ with the time evolution operator being $\mathcal{U}_{0,\tau}=\exp\left(-it\mathcal{H}\right)=\exp\left(-it(H_A+H_B+H_{AB})\right)$.\\
4. At time $\tau$, we separate the systems $A$ and $B$ apart and then we perform the second projective measurement of the conserved quantities of system $A$ and $B$ respectively. The results of the second measurement are $\{\vec{\mathcal{E}}_{m}^A,\vec{\mathcal{E}}_{m}^B,\vec{\mathcal{J}}_{m},\vec{\mathcal{K}}_{m}\}$ and the corresponding conditional probability is given by
\begin{eqnarray}
p_{n\rightarrow m}=\left|\langle m|\mathcal{U}_{0,\tau}|n\rangle\right|^2.
\end{eqnarray}
Due to the weak interact between the systems $A$ and $B$, we then have
\begin{eqnarray}
E_{n}^A+E_n^B&\approx& E_m^A+E_m^B,\\
I_{k,n}^A+I_{k,n}^B&\approx& I_{k,m}^A+I_{k,m}^B,k=1,2,\cdots,M.
\end{eqnarray}
With the two systems initialized in the generalized Gibbs ensemble, the generalized heat exchange between the two systems may be defined as
\begin{eqnarray}
\mathcal{Q}&=&(\vec{\beta}_B-\vec{\beta}_A)\cdot(\vec{\mathcal{E}}_n^A-\vec{\mathcal{E}}_m^A)-\vec{\lambda}\cdot\left(\vec{\mathcal{J}}_n-\vec{\mathcal{J}}_m\right)
-\vec{\alpha}\cdot\left(\vec{\mathcal{K}}_n-\vec{\mathcal{K}}_m\right),\\
&=&(\vec{\beta}_B-\vec{\beta}_A)\cdot(\vec{\mathcal{E}}_m^B-\vec{\mathcal{E}}_n^B)-\vec{\lambda}\cdot\left(\vec{\mathcal{J}}_n-\vec{\mathcal{J}}_m\right)
-\vec{\alpha}\cdot\left(\vec{\mathcal{K}}_n-\vec{\mathcal{K}}_m\right),\\
&=&\Delta\vec{\beta}\cdot(\vec{\mathcal{E}}_n^A-\vec{\mathcal{E}}_m^A)-\vec{\lambda}\cdot\left(\vec{\mathcal{J}}_n-\vec{\mathcal{J}}_m\right)
-\vec{\alpha}\cdot\left(\vec{\mathcal{K}}_n-\vec{\mathcal{K}}_m\right).
\end{eqnarray}
Thus the quantum heat exchange distribution in the GGE is
\begin{eqnarray}\label{hd}
P(Q)&=&\sum_{m,n}p_n(0)p_{n\rightarrow m}\delta\left[Q-\Delta\vec{\beta}\cdot(\vec{\mathcal{E}}_n^A-\vec{\mathcal{E}}_m^A)+\vec{\lambda}\cdot\left(\vec{\mathcal{J}}_n-\vec{\mathcal{J}}_m\right)
+\vec{\alpha}\cdot\left(\vec{\mathcal{K}}_n-\vec{\mathcal{K}}_m\right)\right],\\
&=&\sum_{m,n}p_n(0)\left|\langle m|\mathcal{U}_{0,\tau}|n\rangle\right|^2\delta\left[Q-\Delta\vec{\beta}\cdot(\vec{\mathcal{E}}_n^A-\vec{\mathcal{E}}_m^A)+\vec{\lambda}\cdot\left(\vec{\mathcal{J}}_n-\vec{\mathcal{J}}_m\right)
+\vec{\alpha}\cdot\left(\vec{\mathcal{K}}_n-\vec{\mathcal{K}}_m\right)\right],\\
&=&\sum_{m,n}\frac{e^{-\left(\vec{\beta}_A\cdot\vec{\mathcal{E}}_{n}^A+\vec{\lambda}\cdot\vec{\mathcal{J}}_{n}+\vec{\beta}_B\cdot\vec{\mathcal{E}}_{n}^B+\vec{\alpha}\cdot\vec{\mathcal{K}}_{n}\right)}}{Z_AZ_B}
\left|\langle m|\mathcal{U}_{0,\tau}|n\rangle\right|^2\delta\left[Q-\Delta\vec{\beta}\cdot(\vec{\mathcal{E}}_n^A-\vec{\mathcal{E}}_m^A)+\vec{\lambda}\cdot\left(\vec{\mathcal{J}}_n-\vec{\mathcal{J}}_m\right)
+\vec{\alpha}\cdot\left(\vec{\mathcal{K}}_n-\vec{\mathcal{K}}_m\right)\right].
\end{eqnarray}
The characteristic function for the distribution of heat exchange is given by its Fourier transform,
\begin{eqnarray}
G(u)&=&\int d\mathcal{Q}P(\mathcal{Q})e^{iu\mathcal{Q}},\\
&=&\text{Tr}[e^{-(\mathcal{H}_A+\mathcal{H}_B)}e^{-iu(\mathcal{H}_A+\mathcal{H}_B)}\mathcal{U}_{0,\tau}^{\dagger}e^{iu(\mathcal{H}_A+\mathcal{H}_B)}\mathcal{U}_{0,\tau}].
\end{eqnarray}
Note that the characteristic function for the distribution of heat exchange has similar expression to the quantum decoherence of a single spin coupled to an environment \cite{Yang2017},
which has been deeply related to the partition function of the environment \cite{Wei2012,Wei2014,Wei2015,Peng2015,Wei2017a1,Wei2017b1} and quantum phase transitions in the environments \cite{Sun2006,Zhang2008,Liu2013}.

\subsection{Fluctuation Relations for Heat Exchange in the Generalized Gibbs Ensemble}
\noindent To derive the fluctuation relations for heat exchange. we assume the Hamiltonian $\mathcal{H}=H_A+H_B+H_{AB}$ is time reversal invariant,
\begin{eqnarray}\label{ti}
\Theta^{-1}\mathcal{H}\Theta=\mathcal{H},
\end{eqnarray}
where $\Theta$ is the time reversal operator. Then we have
\begin{eqnarray}
p_n(0)p_{n\rightarrow m}=\frac{e^{-\left(\vec{\beta}_A\cdot\vec{\mathcal{E}}_{n}^A+\vec{\lambda}\cdot\vec{\mathcal{J}}_{n}+\vec{\beta}_B\cdot\vec{\mathcal{E}}_{n}^B+\vec{\alpha}\cdot\vec{\mathcal{K}}_{n}\right)}}{Z_AZ_B}\left|\langle m|\mathcal{U}_{0,\tau}|n\rangle\right|^2.
\end{eqnarray}
Then the probability for the time reversed transitions is
\begin{eqnarray}
p_{\Theta m}(0)p_{\Theta m\rightarrow \Theta n}&=&\frac{e^{-\left(\vec{\beta}_A\cdot\vec{\mathcal{E}}_{m}^A+\vec{\lambda}\cdot\vec{\mathcal{J}}_{m}+\vec{\beta}_B\cdot\vec{\mathcal{E}}_{m}^B+\vec{\alpha}\cdot\vec{\mathcal{K}}_{m}\right)}}{Z_AZ_B}\left|\langle n|\Theta^{-1}\mathcal{U}_{0,\tau}\Theta|m\rangle\right|^2,\\
&=&\frac{e^{-\left(\vec{\beta}_A\cdot\vec{\mathcal{E}}_{m}^A+\vec{\lambda}\cdot\vec{\mathcal{J}}_{m}+\vec{\beta}_B\cdot\vec{\mathcal{E}}_{m}^B+\vec{\alpha}\cdot\vec{\mathcal{K}}_{m}\right)}}{Z_AZ_B}\left|\langle n|\mathcal{U}_{0,\tau}^{\dagger}|m\rangle\right|^2,\label{t1}\\
&=&\frac{e^{-\left(\vec{\beta}_A\cdot\vec{\mathcal{E}}_{m}^A+\vec{\lambda}\cdot\vec{\mathcal{J}}_{m}+\vec{\beta}_B\cdot\vec{\mathcal{E}}_{m}^B+\vec{\alpha}\cdot\vec{\mathcal{K}}_{m}\right)}}{Z_AZ_B}\left|\langle m|\mathcal{U}_{0,\tau}|n\rangle\right|^2. \label{t2}
\end{eqnarray}
From Equation \eqref{t1} to \eqref{t2}, we have made use of the time reversal invariance of the Hamiltonian \eqref{ti}.
Dividing $p_n(0)p_{n\rightarrow m}$ by $p_{\Theta m}(0)p_{\Theta m\rightarrow \Theta n}$, we have
\begin{eqnarray}
\frac{p_n(0)p_{n\rightarrow m}}{p_{\Theta m}(0)p_{\Theta m\rightarrow \Theta n}}&=&e^{-\left(\vec{\beta}_A\cdot\vec{\mathcal{E}}_{n}^A+\vec{\lambda}\cdot\vec{\mathcal{J}}_{n}+\vec{\beta}_B\cdot\vec{\mathcal{E}}_{n}^B+\vec{\alpha}\cdot\vec{\mathcal{K}}_{n}\right)}
e^{\left(\vec{\beta}_A\cdot\vec{\mathcal{E}}_{m}^A+\vec{\lambda}\cdot\vec{\mathcal{J}}_{m}+\vec{\beta}_B\cdot\vec{\mathcal{E}}_{m}^B+\vec{\alpha}\cdot\vec{\mathcal{K}}_{m}\right)},\\
&=&e^{(\vec{\beta}_B-\vec{\beta}_A)\cdot(\vec{\mathcal{E}}_n^A-\vec{\mathcal{E}}_m^A)}e^{-\vec{\lambda}\cdot(\vec{\mathcal{J}}_n-\vec{\mathcal{J}}_m)}e^{-\vec{\alpha}\cdot(\vec{\mathcal{K}}_n-\vec{\mathcal{K}}_m)},\\
&=&e^{\Delta\vec{\beta}\cdot(\vec{\mathcal{E}}_n^A-\vec{\mathcal{E}}_m^A)-\vec{\lambda}\cdot(\vec{\mathcal{J}}_n-\vec{\mathcal{J}}_m)-\vec{\alpha}\cdot(\vec{\mathcal{K}}_n-\vec{\mathcal{K}}_m)},\\
&=&e^{\mathcal{Q}}.
\end{eqnarray}
So
\begin{eqnarray}
P(\mathcal{Q})&=&\sum_{m,n}p_n(0)p_{n\rightarrow m}\delta\left[Q-\Delta\vec{\beta}\cdot(\vec{\mathcal{E}}_n^A-\vec{\mathcal{E}}_m^A)+\vec{\lambda}\cdot\left(\vec{\mathcal{J}}_n-\vec{\mathcal{J}}_m\right)
+\vec{\alpha}\cdot\left(\vec{\mathcal{K}}_n-\vec{\mathcal{K}}_m\right)\right],\\
&=&e^{\mathcal{Q}}\sum_{\Theta m,\Theta n}p_{\Theta m}(0)p_{\Theta m\rightarrow \Theta n}\delta\left[Q-\Delta\vec{\beta}\cdot(\vec{\mathcal{E}}_n^A-\vec{\mathcal{E}}_m^A)+\vec{\lambda}\cdot\left(\vec{\mathcal{J}}_n-\vec{\mathcal{J}}_m\right)
+\vec{\alpha}\cdot\left(\vec{\mathcal{K}}_n-\vec{\mathcal{K}}_m\right)\right],\\
&=&e^{\mathcal{Q}}\sum_{\Theta m,\Theta n}p_{\Theta m}(0)p_{\Theta m\rightarrow \Theta n}\delta\left[Q+\Delta\vec{\beta}\cdot(\vec{\mathcal{E}}_m^A-\vec{\mathcal{E}}_n^A)+\vec{\lambda}\cdot\left(\vec{\mathcal{J}}_m-\vec{\mathcal{J}}_n\right)
+\vec{\alpha}\cdot\left(\vec{\mathcal{K}}_m-\vec{\mathcal{K}}_n\right)\right],\\
&=&e^{\mathcal{Q}}P(-\mathcal{Q}).
\end{eqnarray}
We thus derived the exchange fluctuation relations in the GGE,
\begin{eqnarray}\label{ga}
\frac{P(\mathcal{Q})}{P(-\mathcal{Q})}=e^{\mathcal{Q}}.
\end{eqnarray}
This is a generalization of the heat exchange fluctuation theorem derive by Jarzynksi and W\'{o}jcik \cite{Jarzynski2004} into very general family of initial conditions.
From Equation \eqref{ga}, we get
\begin{eqnarray}\label{gc}
\left\langle e^{-\mathcal{Q}}\right\rangle=\int d\mathcal{Q}P(\mathcal{Q})e^{-\mathcal{Q}}=\int d\mathcal{Q}P(-\mathcal{Q})=1.
\end{eqnarray}
If the two quantum systems $A$ and $B$ have no conserved quantity except the energy of the system, then the generalized heat exchange reduces to
\begin{eqnarray}
\mathcal{Q}=\Delta\beta(E_n^A-E_m^A)=(\beta_B-\beta_A)(E_n^A-E_m^A)=\Delta\beta Q.
\end{eqnarray}
Thus Equation \eqref{ga} reduces to Equation\eqref{fa}.

\section{Relations between Heat Exchange and R\'{e}nyi Divergences in the Generalized Gibbs Ensemble}
Recently the author and its collaborator found that the dissipated work (work minus the free energy difference) in a non-equilibrium process is related to the Renyi divergences
between microscopic states in the forward and reversed dynamics under very general family of initial conditions \cite{Wei2017a,Wei2017b,Wei2017c} and this relation has recently been tested in a superconducting qubit system \cite{Fan2017}.
In the section, we generalize the relations between heat exchange and the Renyi divergences \cite{Wei2017d} to quantum systems whose equilibrium states are described by GGE. With the heat distribution function \eqref{hd},
we have the generating function of heat exchange between $A$ and $B$
\begin{eqnarray}
\left\langle \Big(e^{-\mathcal{Q}}\Big)^z\right\rangle&=&\int d\mathcal{Q}P(\mathcal{Q})e^{-z \mathcal{Q}},\\
&=&\sum_{m,n}\frac{e^{-\left(\vec{\beta}_A\cdot\vec{\mathcal{E}}_{n}^A+\vec{\lambda}\cdot\vec{\mathcal{J}}_{n}+\vec{\beta}_B\cdot\vec{\mathcal{E}}_{n}^B+\vec{\alpha}\cdot\vec{\mathcal{K}}_{n}\right)}}{Z_AZ_B}\left|\langle m|\mathcal{U}_{0,\tau}|n\rangle\right|^2
e^{-z(\vec{\beta}_B-\vec{\beta}_A)\cdot(\vec{\mathcal{E}}_n^A-\vec{\mathcal{E}}_m^A)+z\vec{\lambda}\cdot(\vec{\mathcal{J}}_n-\vec{\mathcal{J}}_m)+z\vec{\alpha}\cdot(\vec{\mathcal{K}}_n-\vec{\mathcal{K}}_m)},\\
&=&\sum_{m,n}\frac{e^{-\left(\vec{\beta}_A\cdot\vec{\mathcal{E}}_{n}^A+\vec{\lambda}\cdot\vec{\mathcal{J}}_{n}+\vec{\beta}_B\cdot\vec{\mathcal{E}}_{n}^B+\vec{\alpha}\cdot\vec{\mathcal{K}}_{n}\right)}}{Z_AZ_B}\left|\langle m|\mathcal{U}_{0,\tau}|n\rangle\right|^2
e^{-z\vec{\beta}_B\cdot(\vec{\mathcal{E}}_m^B-\vec{\mathcal{E}}_n^B)}
e^{z\vec{\beta}_A\cdot(\vec{\mathcal{E}}_n^A-\vec{\mathcal{E}}_m^A)}e^{z\vec{\lambda}\cdot(\vec{\mathcal{J}}_n-\vec{\mathcal{J}}_m)+z\vec{\alpha}\cdot(\vec{\mathcal{K}}_n-\vec{\mathcal{K}}_m)},\\
&=&\frac{1}{Z_AZ_B}\sum_{m,n}\langle m|\mathcal{U}_{0,\tau}|n\rangle \langle n|\mathcal{U}_{0,\tau}^{\dagger}|m\rangle e^{-(1-z)\left(\vec{\beta}_A\cdot\vec{\mathcal{E}}_{n}^A+\vec{\lambda}\cdot\vec{\mathcal{J}}_{n}+\vec{\beta}_B\cdot\vec{\mathcal{E}}_{n}^B+\vec{\alpha}\cdot\vec{\mathcal{K}}_{n}\right)}
e^{-z\left(\vec{\beta}_A\cdot\vec{\mathcal{E}}_m^A+\vec{\lambda}\cdot\vec{\mathcal{J}}_m+\vec{\beta}_B\cdot\vec{\mathcal{E}}_m^B+\vec{\alpha}\cdot\vec{\mathcal{K}}_m\right)},\\
&=&\frac{1}{Z_AZ_B}\sum_{m,n}\langle m|\mathcal{U}_{0,\tau}e^{-(1-z)(\mathcal{H}_A+\mathcal{H}_B)}|n\rangle \langle n|\mathcal{U}_{0,\tau}^{\dagger}e^{-z(\mathcal{H}_A+\mathcal{H}_B)}|m\rangle,\\
&=&\frac{1}{Z_AZ_B}\text{Tr}\left[\mathcal{U}_{0,\tau}e^{-(1-z)(\mathcal{H}_A+\mathcal{H}_B)}\mathcal{U}_{0,\tau}^{\dagger}
e^{-z(\mathcal{H}_A+\mathcal{H}_B)} \right],\\
&=&\frac{1}{Z_AZ_B}\text{Tr}\left[\left(\mathcal{U}_{0,\tau}e^{-(\mathcal{H}_A+\mathcal{H}_B)}\mathcal{U}_{0,\tau}^{\dagger}\right)^{1-z}\left(e^{-(\mathcal{H}_A+\mathcal{H}_B)}\right)^z \right],\\
&=&\text{Tr}\left[\left(\mathcal{U}_{0,\tau}\rho(0)\mathcal{U}_{0,\tau}^{\dagger}\right)^{1-z}\rho(0)^z \right],\\
&=&\text{Tr}\left[\rho(\tau)^{1-z}\rho(0)^z \right],\\
&=&e^{(z-1)S_z\left(\rho(0)||\rho(\tau)\right)}.
\end{eqnarray}
Here we have made use of the shorthand notation $\mathcal{H}_A=\beta_0^A H_A+\sum_{i=1}^M\beta_{i}^AI_{i}+\sum_{i=1}^{M'}\lambda_{i}J_{i}$ and $\mathcal{H}_B=\beta_0^B H_B+\sum_{i=1}^M\beta_{i}^BI_{i}+\sum_{i=1}^{N}\alpha_{i}K_{i}$.
We thus proved that the generating function of heat exchange is related to the R\'{e}nyi divergences between the initial equilibrium state and final out of equilibrium state in the generalized Gibbs ensemble,
\begin{eqnarray}\label{gb}
\left\langle \Big(e^{-\mathcal{Q}}\Big)^z\right\rangle=e^{(z-1)S_z\left(\rho(0)||\rho(\tau)\right)}.
\end{eqnarray}
Finally we make several remarks on Equation \eqref{gb}:\\
1. As $z$ is an arbitrary real number in Equation \eqref{gb}, we set $z=1$ and yield
\begin{eqnarray}
\left\langle e^{-\mathcal{Q}}\right\rangle&=&1.
\end{eqnarray}
2. From Equation \eqref{gb}, the average heat exchange in the GGE is
\begin{eqnarray}
\langle \mathcal{Q}\rangle&=&D\left(\rho(\tau)||\rho(0)\right),
\end{eqnarray}
where the right hand side is the relative entropy \cite{relativeentropy} between microscopic states $\rho(\tau)$ and $\rho(0)$,
which are the final out of equilibrium state of the total system and the initial equilibrium state of the total system, respectively.
Furthermore, the other moments of the heat exchange in the GGE is
\begin{eqnarray}
\langle\mathcal{Q}^n\rangle&=&\text{Tr}\left[\rho(\tau)\mathcal{T}_n\left(\ln[\rho(\tau)]-\ln[\rho(0)]\right)^n\right],
\end{eqnarray}
where $n=1,2,3,\cdots$ and $\mathcal{T}_n$ is an ordering operator so that in the binomial expansion of $\left(\ln[\rho(\tau)]-\ln[\rho(0)]\right)^n$, $\ln[\rho(\tau)]$ always sits on the left of $\ln[\rho(0)]$.\\
3. It is known in information theory that Renyi divergences is a measure of the distinguishability between two states \cite{Renyi1961,Erven2014,Beigi2013,Lennert2013}
and therefore the Renyi divergence in equation \eqref{gb} implies that heat exchange is a consequence of out of equilibrium dynamics of two systems. \\
4. Equation \eqref{gb} is true for any contact time $\tau$ between two systems and this is a result of unitarity in dynamics of quantum systems.\\
5. If the two quantum systems $A$ and $B$ have no conserved quantity except their energies, then the generalized heat reduces to
\begin{eqnarray}
\mathcal{Q}=\Delta\beta(E_n^A-E_m^A)=(\beta_B-\beta_A)(E_n^A-E_m^A)=\Delta\beta Q.
\end{eqnarray}
Thus Equation \eqref{gb} reduces to Equation\eqref{fb}.\\
6. The heat exchange in the Gibbs ensemble may be extracted from the Ramsey interference of a single spin \cite{Goold2014}. It is conceivable that the same method may be adopted to measure
the heat exchange between quantum systems which are described by the generalized Gibbs ensemble. Therefore the central results in this work, the fluctuation relations for heat exchange in the GGE \eqref{ga}
and the relation between heat exchange and the Renyi divergences \eqref{gb} may be verified experimentally.

\section{Conclusions}
We have investigated the heat exchange between two quantum systems initialized in the equilibrium states described by the generalized Gibbs ensemble.
We have found that statistics of heat exchange satisfies fluctuation relations for quantum systems initialized in equilibrium states described by the generalized Gibbs ensemble at different generalized temperatures if we adopt the concept
of generalized heat exchange. This fluctuation relation is a generalization of the heat exchange fluctuation relation by Jarzynski and W\'{o}jcik to quantum systems described by generalized Gibbs ensemble. 
Moreover, we have extended the connections between heat exchange and R\'{e}nyi divergences to quantum systems initialized in the generalized Gibbs ensemble.
These relations are applicable for quantum systems with conserved quantities and are universally valid for quantum systems in the integrable and chaotic regimes.

\begin{acknowledgements}
This work was supported by the National Natural Science Foundation of China (Grant Number 11604220) and the Startup Fund of Shenzhen University under (Grant number 2016018).
\end{acknowledgements}


\begin{references}
\bibitem{Jarzynski2004}
C. Jarzynski and D. K. W\'{o}jcik, Classical and Quantum Fluctuation Theorems for Heat Exchange, Phys. Rev. Lett. \textbf{92}, 230601 (2004).



\bibitem{Wei2017d}
B. B. Wei, Relations between Heat Exchange and the R\'{e}nyi Divergences, arXiv:1711.05383 (2017).



\bibitem{Renyi1961}
A. R\'{e}nyi, in \emph{Proceedings of the Fourth Berkeley Symposium
on Mathematical Statistics and Probability} (University of
California Press Press, 1961), pp. 547-561.

\bibitem{Erven2014}
T. Van Erven and P Harremos, R\'{e}nyi Divergence and Kullback-Leibler Divergence, IEEE Transactions on Information Theory, \textbf{60}, 3797 (2014).


\bibitem{Beigi2013}
S. Beigi, Sandwiched R\'{e}nyi divergence satisfies data processing inequality, J. Math. Phys. \textbf{54}, 122202 (2013).

\bibitem{Lennert2013}
M. M\"{u}ller-Lennert, F. Dupuis, O. Szehr, S. Fehr, M. Tomamichel, On quantum R\'{e}nyi entropies: A new generalization and some properties, J. Math. Phys. \textbf{54}, 122203 (2013).


\bibitem{Jar2011}
C. Jarzynski, Equalities and inequalities: Irreversibility and the second law of thermodynamics at the nanoscale, Annu. Rev. Condens. Matter Phys. \textbf{2}, 329 (2011).

\bibitem{RMP2011}
M. Campisi,  P. Hanggi and P. Talkner, Colloquium: Quantum fluctuation relations: Foundations and applications, Rev. Mod. Phys. \textbf{83}, 771 (2011).




\bibitem{GGE0}
T. Kinoshita, T. Wenger, D. S. Weiss, A quantum Newton's cradle, Nature \textbf{440}, 900 (2006).


\bibitem{GGE1}
M. Rigol, V. Dunjko, V. Yurovsky, M. Olshanii, Relaxation in a Completely Integrable Many-Body Quantum System: An \emph{Ab Initio} Study of the Dynamics of the Highly Excited States of 1D Lattice Hard-Core Bosons, Phys. Rev. Lett. \textbf{98}, 050405 (2007).


\bibitem{GGE2}
P. Calabrese, F. H. L. Essler, M. Fagotti, Quantum Quench in the Transverse Field Ising Chain, Phys. Rev. Lett. \textbf{106}, 227203 (2011).



\bibitem{GGE3}
M. Gring et al., Relaxation and Pre-thermalization in an Isolated Quantum System, Science 337, 1318 (2012).

\bibitem{GGE4}
J. P. Ronzheimer, \emph{et al.} Expansion Dynamics of Interacting Bosons in Homogeneous Lattices in One and Two Dimensions, Phys. Rev. Lett. \textbf{110}, 205301 (2013).

\bibitem{GGE5}
J. S. Caux, F. H. Essler, Time Evolution of Local Observables After Quenching to an Integrable Model, Phys. Rev. Lett. \textbf{110}, 257203 (2013).


\bibitem{GGE6}
L. Vidmar, \emph{et. al.} Dynamical Quasicondensation of Hard-Core Bosons at Finite Momenta, Phys. Rev. Lett. \textbf{115}, 175301 (2015).

\bibitem{GGE7}
L. Vidmar, D. Iyer and M. Rigol, Emergent Eigenstate Solution to Quantum Dynamics Far from Equilibrium, Phys. Rev. X \textbf{7}, 021012 (2017).



\bibitem{GGEreview2016}
C. Gogolin and J. Eisert, Equilibration, thermalisation, and the emergence of statistical mechanics in closed quantum systems, Rep. Prog. Phys. \textbf{79}, 056001 (2016).


\bibitem{GGEreview2016b}
L. Vidmar and M. Rigol, Generalized Gibbs ensemble in integrable lattice models, J. Stat. Mech. Theor. Exp. (2016) 064007.



\bibitem{GGEexp}
T. Langen, \emph{et al.} Experimental observation of a generalized Gibbs ensemble, Science, \textbf{348}, 207 (2015).



\bibitem{Reichl1987}
L. E. Reichl, \emph{A Modern Course in Statistical Physics} (Edward Arnold, Austin, TX, 1987).



\bibitem{Jaynes1957a}
E. T. Jaynes, Information Theory and Statistical Mechanics, Phys. Rev. \textbf{106}, 620 (1957).

\bibitem{Jaynes1957b}
E. T. Jaynes, Information Theory and Statistical Mechanics. II, Phys. Rev. \textbf{108}, 171 (1957).






\bibitem{Yang2017}
W. Yang, W. L. Ma and R. B. Liu, Quantum many-body theory for electron spin decoherence in nanoscale nuclear spin baths, Rep. Prog. Phys. 80, 016001 (2017).


\bibitem{Wei2012}
B. B. Wei and R. B. Liu, Lee-Yang zeros and critical times in decoherence of a probe spin coupled to a bath, Phys. Rev. Lett. \textbf{109}, 185701 (2012).

\bibitem{Wei2014}
B. B. Wei, S. W. Chen, H. C. Po and R. B. Liu, Phase transitions in the complex plane of a physical parameter, Sci. Rep. \textbf{4}, 5202 (2014).

\bibitem{Wei2015}
B. B. Wei, Z. F. Jiang and R. B. Liu, Thermodynamic holography, Sci. Rep. \textbf{5}, 15077 (2015).

\bibitem{Peng2015}
X. H. Peng, H. Zhou, B. B. Wei, J. Y. Cui, J. F. Du and R. B. Liu, Experimental observation of Lee-Yang zeros, Phys. Rev. Lett. \textbf{114}, 010601 (2015).


\bibitem{Wei2017a1}
B. B. Wei, Probing Yang-Lee edge singularity by central spin decoherence, New J. Phys. \textbf{19}, 083009 (2017).

\bibitem{Wei2017b1}
B. B. Wei, Probing conformal invariant of non-unitary two-dimensional system by central spin decoherence, arXiv:1708.03241 (2017).



\bibitem{Sun2006}
H. T. Quan, Z. Song, X. F. Liu, P. Zanardi, and C. P. Sun, Decay of Loschmidt echo enhanced by quantum criticality, Phys.
Rev. Lett. \textbf{96}, 140604 (2006).

\bibitem{Zhang2008}
J. Zhang, X. Peng, N. Rajendran, and D. Suter, Detection of quantum critical points by a probe qubit, Phys. Rev. Lett. \textbf{100}, 100501 (2008).


\bibitem{Liu2013}
S. W. Chen, Z. F. Jiang and R. B. Liu, Quantum criticality at high temperature revealed by spin echo, New J. Phys. \textbf{15}, 043032 (2013).


\bibitem{Wei2017a}
B. B. Wei and M. B. Plenio, Relations between dissipated work in non-equilibrium process and the family of R\'{e}nyi divergences, New J. Phys. \textbf{19}, 023002(2017).


\bibitem{Wei2017b}
B. B. Wei, Links between dissipation and Renyi divergences in the $\mathcal{PT}$-symmetric quantum mechanics, arXiv:1710.06059 (2017).


\bibitem{Wei2017c}
B. B. Wei, Dissipation in the Generalized Gibbs Ensemble,  arXiv:1711.03855 (2017).


\bibitem{Fan2017}
X. Y. Guo, et al. Demonstration of irreversibility and dissipation relation of thermodynamics with a superconducting qubit, arXiv:
1710.10234 (2017).








\bibitem{relativeentropy}
A. Wehrl, General properties of entropy, Rev. Mod. Phys. \textbf{50}, 221 (1978).




\bibitem{Goold2014}
J. Goold, U. Poschinger and K. Modi, Measuring heat exchange of a quantum process, Phys. Rev. E \textbf{90}, 020101 (2014).











\end{references}
\end{document}